\documentclass[journal]{IEEEtran}
\usepackage{amssymb}
\usepackage{bm}
\usepackage{algorithm} 
\usepackage[noend]{algpseudocode}
\usepackage{amstext}
\usepackage{amsmath}
\usepackage{graphicx}
\usepackage{booktabs}
\usepackage{array}
\ifCLASSINFOpdf
\else

\fi

\begin{document}
\title{Does Explicit Prediction Matter in Deep Reinforcement Learning-Based Energy Management?}

\author{Zhaoming~Qin,~Huaying~Zhang,~Yuzhou~Zhao,~Hong~Xie,~and~Junwei~Cao,~\IEEEmembership{Senior~Member,~IEEE}
\thanks{This work was supported by Science and Technology Project of China Southern Power Grid (090000KK52190169/SZKJXM2019669).}
\thanks{Z. Qin is with the Department of Automation, Tsinghua University, Beijing, P. R. China.}
\thanks{H. Zhang, Y. Zhao and H. Xie are with New Smart City High-quality Power Supply Joint Laboratory of China Southern Power Grid (Shenzhen Power Supply CO., LTD).}
\thanks{J. Cao is with Beijing National Research Center for Information Science and Technology, Tsinghua University, Beijing, P. R. China.}
\thanks{Corresponding author: Junwei Cao, email: jcao@tsinghua.edu.cn.}}

\markboth{IEEE International Conference on Energy Internet, September~2021}%
{Shell \MakeLowercase{\textit{et al.}}: Bare Demo of IEEEtran.cls for IEEE Journals}

\maketitle

\begin{abstract}
As a model-free optimization and decision-making method, deep reinforcement learning (DRL) has been widely applied to the filed of energy management in energy Internet. 
While, some DRL-based energy management schemes also incorporate the prediction module used by the traditional model-based methods, which seems to be unnecessary and even adverse.
In this work, we implement the standard energy management scheme with prediction using supervised learning and DRL, and the counterpart without prediction using end-to-end DRL.
Then, these two schemes are compared in the unified energy management framework. 
The simulation results demonstrate that the energy management scheme without prediction is superior over the scheme with prediction.
This work intends to rectify the misuse of DRL methods in the field of energy management.
\end{abstract}


\IEEEpeerreviewmaketitle

\section{Introduction}

\IEEEPARstart{A}{s} alternative to conventional fossil fuels, there have been large-scale integration of 
the renewable energy sources (RESs) including solar power and wind power into power system \cite{RES,TSE_Hua}. 
Although RESs have advantages including sustainable and environmental friendly,
it is intractable to conduct energy management with the penetration of high-proportional RESs 
due to the uncertainty and stochasticity of renewable generation output \cite{Penetration}.
Moreover, the challenges for energy management are further exacerbated 
by the varying power demands and fluctuating electricity prices \cite{demand,price}.
Therefore, it is of great importance to develop the advanced energy management scheme to accommodates various disturbances from RESs, power demands and electricity prices.

Tremendous research effort has been dedicated in developing the \textit{model-based} energy management schemes \cite{hourahead,MPCTII,EMPC,MPCDSM}. 
A typical model-based approach is model predictive control (MPC), 
in which control signals are decided by solving an optimization problem with a finite time horizon, following a receding horizon approach.
The formulated optimization problem generally relies on the access to full knowledge
of the system model and parameters. Put differently, the optimal energy management scheduling is estimated using forecasted
exogenous parameters, including the power demands, electricity prices and weather-dependent PV production.
As a result, the performance of the consequent energy management schemes is significantly dependent to the accuracy of the employed
system model and the forecasting method. Therefore, massive advanced predictive models and approaches have been developed \cite{TSGyuchao,LSTM}.
A novel hybrid modeling method using both deep neural networks (DNNs) and stochastic differential equations is proposed to To obtain accurate
power models of photovoltaic panels and loads in \cite{TSGyuchao}. 
The long short-term memory recurrent neural network (RNN) is employed to address the short-term residential load forecasting issue in \cite{LSTM}.

By contrast, the model-free energy management schemes do not require the explicit system model and the predictive exogenous parameters, 
regarded as a potential alternative to model-based schemes \cite{DNNRL,ANNRL}.
For example, the model-free reinforcement learning (RL) can 
gradually learn the optimal or near-optimal strategies by utilizing
experiences collected from massive interactions with the environment,
without a priori knowledge of the environment \cite{APEN_Hua}. 
Moreover, with the booming development of deep learning (DL) technologies, 
the deep reinforcement learning (DRL) has attracted great attention \cite{DQN}.
The DRL can be viewed as the combination of DL and RL.
The powerful representation capability of DNNs enables DRL to address the continuous and high-dimensional state spaces and action spaces \cite{DRLSurvey}.
An energy management algorithm based on deep deterministic policy gradient (DDPG) is proposed 
to minimize the energy cost of smart home in \cite{SHEM20}.
Authors in \cite{TSGQin} develop an vectorized DRL algorithm based on advantage actor-critic (A2C) to reduce the operation cost and improve user experience without some users' private information.

Although the DRL-based methods do not rely on the predictive models and parameters, some works still integrate the forecasting methods into model-free DRL, such as \cite{XuXu,HVACpredict}.
Authors in \cite{XuXu} use feedforward DNNs to predict the future electricity prices which are served as the part of observation in DRL.
Similarly, authors in \cite{HVACpredict} establish a price forecasting model using multilayer perceptron (MLP) used for the decision-making of DDPG algorithm.
Prediction is indeed a dimensionality reduction processing of the original information, in this sense, the information received by the agent of DRL is not complete.
Consequently, the powerful feature extraction capability cannot fully utilized.
Despite the essential role of prediction in model-based methods, adding prediction to model-free methods may undermine the control effect.

In this work, we investigate the performance comparison of energy management schemes with and without explicit prediction.
First, we formulate the general energy management problem of a microgrid as a Markov decision process (MDP).
Second, we realize the energy management scheme with prediction, by training the forecasting models with SL and the policy with DRL respectively.
Third, we implement the energy management scheme without prediction, by training the end-to-end policy network consisting of MLP and RNN.
Finally, we conduct the simulation experiments to compare the effects of these two schemes.
The main contributions of this paper can be summarized as follows.
\begin{itemize}
  \item We investigate the effects of prediction in the DRL-based energy management scheme. 
  To the best of our knowledge, this is the first paper to make a rigorous comparison between the DRL-based scheme with and without prediction.
  \item We establish the unified energy management framework under which the comparison between DRL-based scheme with and without prediction can be conducted fairly.
  \item Simulation results demonstrate that the DRL-based scheme without prediction outperforms over the scheme with prediction.
  Moreover, we intuitively explain how the prediction undermines the control effect of DRL.
\end{itemize}

\section{Problem Formulation}
\subsection{System Decription}
\begin{figure}[htbp]
  \centering{\includegraphics[width=.45\textwidth]{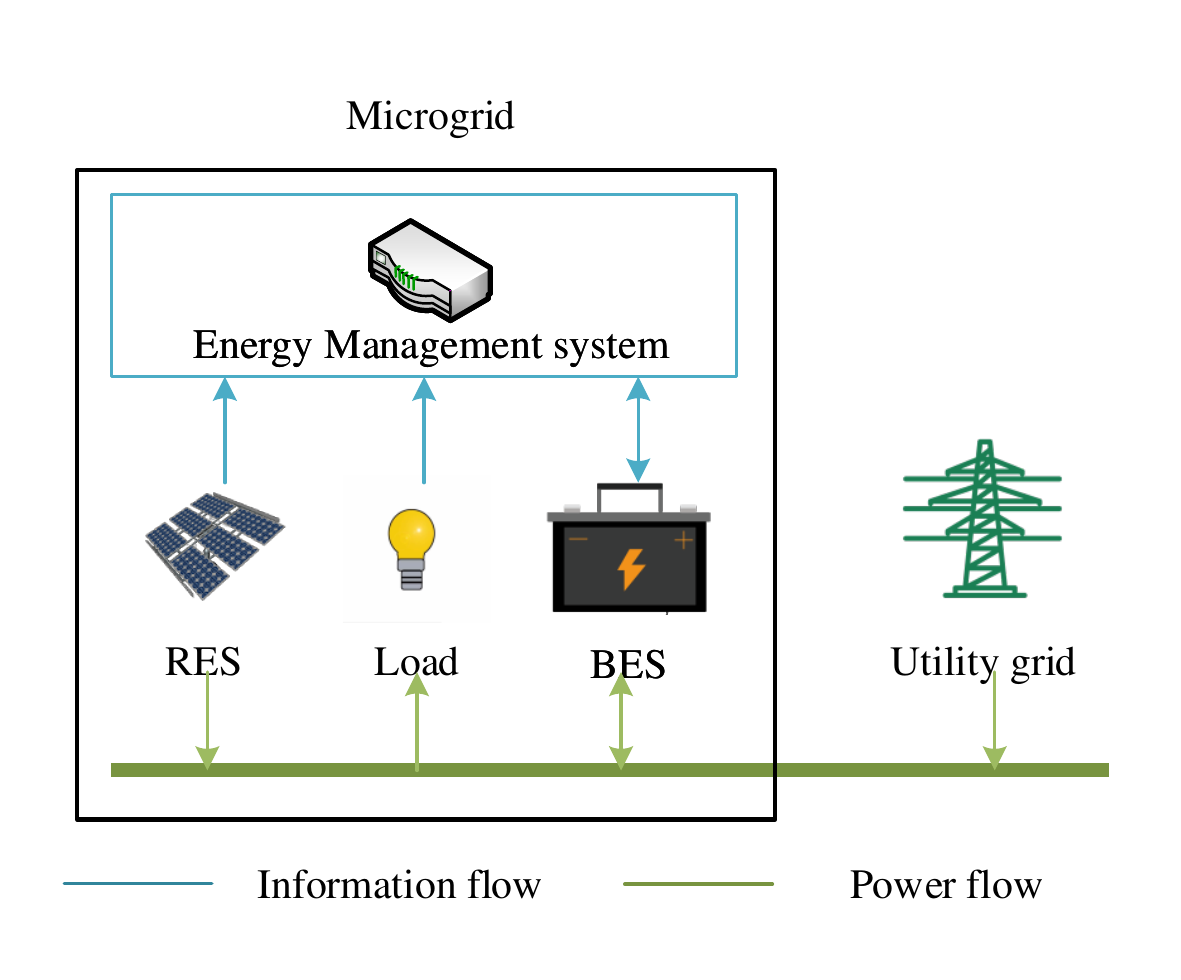}}
  \caption{Illustration of considered microgrid.}
  \label{fig:system}
\end{figure}

In this work, wo consider a general energy management problem of a microgrid.
As shown in Fig. \ref{fig:system}, the microgrid is comprised of RESs, non-adjustable loads, battery energy storage devices (BESs) and energy management system (EMS). 
The RESs could be solar panels and wind generators. The power demands of non-adjustable loads must be satisfied completely without delay.
We suppose that the microgrid hourly operates in discrete time, i.e., $t\in\{0,1,\dots,T\}$ where $T$ is the time horizon.
Each time step begins at the beginning of the current hour and expires at the beginning of the next hour. 
For example, the period from 0:00 to 1:00 is time step 1, the period from 1:00 to 2:00 is time step 2, and so on. 
Moreover, the electricity price is announced hourly by the utility grid.

At the beginning of each hour, the EMS observes the renewable generation output and power demand during last hour, receives current state of charge (SOC) from BESs and hour-ahead electricity price from the utility grid.
Then, the EMS determines the charging/discharging power of BESs. After the decision of EMS, if energy shortage occurs during this hour, the microgrid will purchase appropriate energy from the utility gird; while the excess energy will be abandoned.

\subsection{Markov Decision Process Formulation}
In this work, the energy management scheme is formulated as a MDP.
A general MDP can be described as a tuple $\left(\mathcal{S},\mathcal{A},\mathcal{P},{R}\right)$, 
where $\mathcal{S},\mathcal{A},\mathcal{P},{R}$ are the state space, action space, transition dynamics and reward function. 
At each time step $t$, the agent observe a state $\mathbf{s}_t$ 
from the state space $\mathcal{S}$, and selects an action $a_t$ from action space $\mathcal{A}$.
After performing action $a_t$, state $\mathbf{s}_t$ transitions to state $\mathbf{s+1}_t$ 
with probability distribution $\mathcal{P}(\mathbf{s}_t,a_t)$. Additionally, the agent receives a scalar
reward $r_t=R(\mathbf{s}_t,a_t)$. The goal of the MDP is to maximize the cumulative discount reward $R_0=\sum_{t=0}^T\gamma^tr_t$.
In the remainder of the section, the state, action, dynamics and reward will be specified.
\subsubsection{State}
At each time step $t$, the state available to EMS includes the renewable generation output and power demand at last time step, the current SOC of BESs and the hour-ahead electricity price.
\begin{IEEEeqnarray}{rCl}
  \bm{s}_t = \left[b_t, g_{t-1},d_{t-1},p_t\right],
\end{IEEEeqnarray}
where $b_t$, $g_{t-1}$, $d_{t-1}$ and $p_t$ are the SOC of BESs,
renewable generation output, nonshiftable power demand and electricity price, respectively. 

\subsubsection{Action}
The action $a_t$ denotes the charging/discharging power at time step $t$, constrained by
\begin{IEEEeqnarray}{rCl}
  -d_{max} \leq a_{t} \leq c_{max},
\end{IEEEeqnarray}
where $d_{max}$ and $c_{max}$ are the maximum discharging power and charging power, respectively.
\subsubsection{Dynamics}
Normally, the dynamics of generation output, power demand and electricity price are difficult to describe precisely.
In this work, the real historical data is directly utilized, including power data \cite{Database} and price data \cite{ENGIE}. 

The dynamics of SOC is presented as follows
\begin{IEEEeqnarray}{rCl}
  b_{t+1} = f\left(b_{t}, a_t\right),
\end{IEEEeqnarray}
where $f(\cdot)$ denote the transition function of SOC with respect to current SOC and charging/discharging power.

The power balance is guaranteed by purchasing energy from the utility grid.
\begin{IEEEeqnarray}{rCl}
  e_t = 
  \begin{cases}
    d_{t}  - g_{t} + b_t,&\mbox{if } d_{t}  - g_{t} + b_t>0,\\
    0,&\mbox{otherwise},\\
  \end{cases}
\end{IEEEeqnarray}
where $e_t$ is the power drawn from the utility grid at time step $t$.

\subsubsection{Reward}
The negative reward at each time step can be divided into two parts: the energy transaction with utility grid and the degradation cost of BESs. Thus, the reward is presented as follows,
\begin{IEEEeqnarray}{rCl}
  r_{t} = -p_{t}\cdot e_t - g\left(b_{t}, a_t\right),
\end{IEEEeqnarray}
where $g(\cdot)$ is the degradation cost function of BESs with respect to current SOC and charging/discharging power.

\section{DRL-Based Energy Management Scheme}
In this section, we present the DRL-based scheme with and without prediction.
First, SL is applied to train a RNN to conduct the prediction.
Then, the DRL-based scheme with the prediction module is realized.
Finally, the end-to-end DRL-based scheme without prediction is proposed. 
\subsection{SL for Prediction}
The target of SL is to learn a function $f$ parameterized by $\phi$
such that $y=f(x;\varphi)$.
Here, $x$ and $y$ denote the input and the label, respectively. 
Under our scenario, we intend to predict the future renewable generation output, power demand and hour-ahead price, which belongs to time series prediction problem.
Considering the outperformance of RNNs with the processing of temporal sequence, in this work, gated recurrent units (GRUs), a gating mechanism in RNNs, 
are employed to represent the forecasting models.

The training process of SL is shown in Algorithm \ref{alg:SL}.
The input $x_t$ could be generation output $g_t$, power demand $d_t$ and electricity price $p_t$. 
The GRUs are trained with back-propagation such that the mean square error (MSE) between the outputs of GRUs and the target values is minimized.   
\begin{algorithm}
  \caption{SL for $k$-step prediction} \label{alg:SL}
  \begin{algorithmic}
      \State Initialize parameter $\varphi_k$
      \For{epoch $=1$ to $N$}
        \For{time step $t=0$ to $T$}
          \State $\hat{x}_{t+k},h_{t+1} = GRU(x_{t},h_{t};\varphi_k)$
          \State $d\varphi_k \leftarrow d\varphi_k + \nabla_{\varphi_k}\left(\hat{x}_{t+k}-x_{t+k}\right)^2$
        \EndFor
        \State Perform update of $\varphi_k$ using $d\varphi_k$
      \EndFor
  \end{algorithmic}
\end{algorithm}

\subsection{DRL-Based Scheme with Prediction}
Given the trained forecasting models, in this subsection, the DRL-based scheme with prediction is presented.

First, the observation of DRL is the concatenation of state and prediction as follows, 
\begin{IEEEeqnarray}{rCl}
  \bm{o}_t = \bm{s}_t\cup \left[\hat{g}_{t},\hat{d}_{t},\hat{p}_{t+1},\dots,\hat{g}_{t+k},\hat{d}_{t+k},\hat{p}_{t+k}\right],
\end{IEEEeqnarray}
where the expanded part is predicted by the forecasting models.
\begin{figure}[htbp]
  \centering{\includegraphics[width=.45\textwidth]{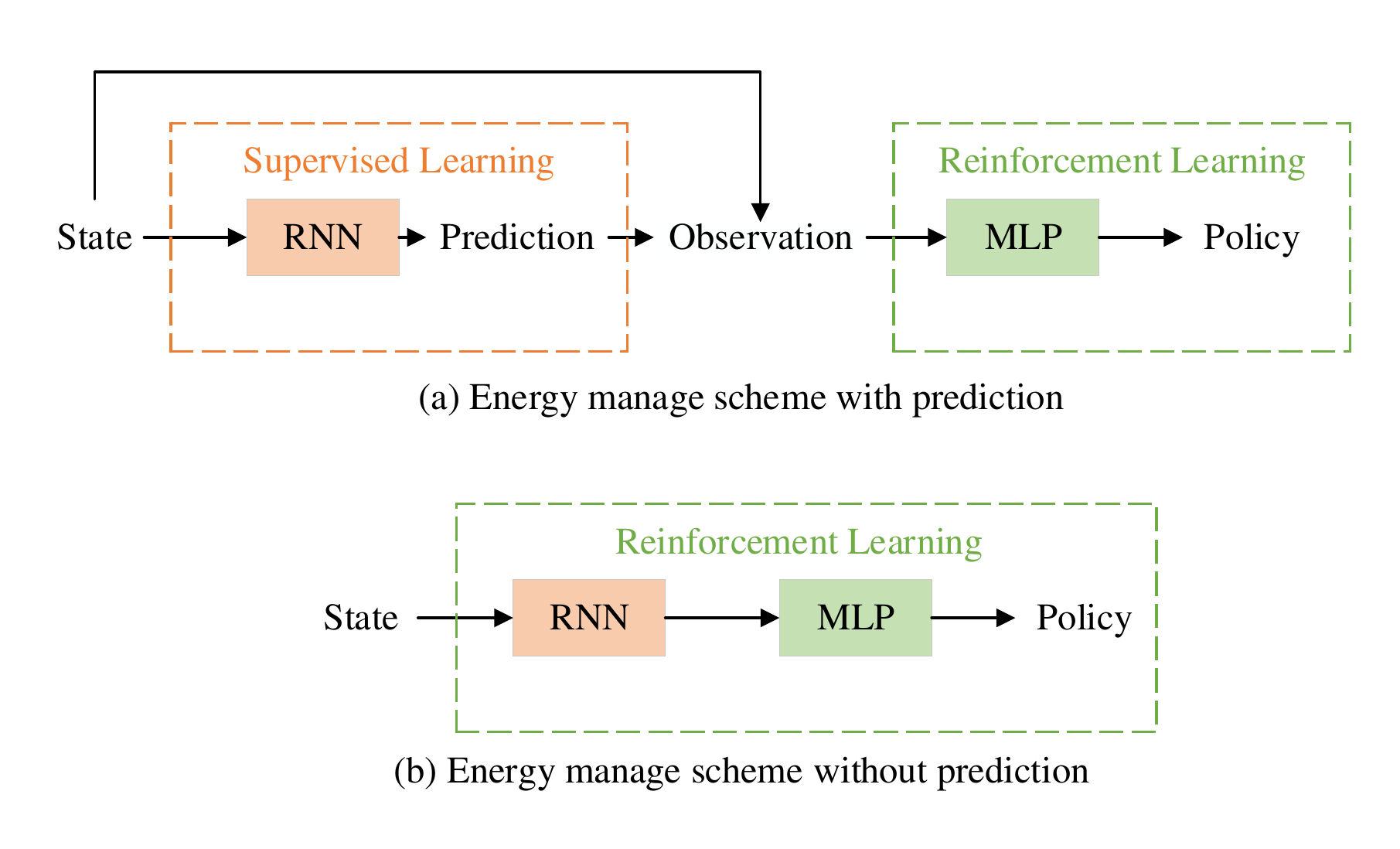}}
  \caption{Overview of energy management schemes.}
  \label{fig:network}
\end{figure}

Then, the policy of DRL is trained by PPO algorithm \cite{PPO} which maintains the actor network $\pi\left(a_t|\mathbf{o}_t;\theta\right)$ parameterized by $\theta$ 
and the critic network $V(\mathbf{o}_t;\phi)$ parameterized by $\phi$. The parameters of actor are updated by minimizing following loss function,
\begin{IEEEeqnarray}{rCl}\label{La}
  \mathcal{L}_a(\theta) = \mathbb{E}_{t}\left[ \min \left(w_t\hat{A}_t,\text{clip}\left(w_t,1-\epsilon,1+\epsilon\right)\hat{A}_t\right)\right].
\end{IEEEeqnarray}
In (\ref{La}), $\epsilon$ is a hyperparameter controlling the range of clipped objective; $\hat{A}_t$ denotes the advantage estimated by general advantage estimation (GAE) method \cite{GAE}
\begin{IEEEeqnarray}{rCl}
  \hat{A}_t = \sum_{t'=t}^{T}\left(\gamma\lambda\right)^{t'-t}\left(-V(\mathbf{o}_{t'};\phi)+r_t+\gamma V\left(\mathbf{o}_{t'+1};\phi\right)\right),
\end{IEEEeqnarray}
where $\gamma$ is a hyperparameter employed to control the trade-off between variance and bias of the estimate.
In (\ref{La}), $w_t$ is the probability ratio defined as  
\begin{IEEEeqnarray}{rCl}
  w_t=\frac{\pi(a_t|\mathbf{o}_t;\theta)}{\pi(a_t|\mathbf{o}_t;\theta_{\text{old}})},
\end{IEEEeqnarray}
where $\theta_{\text{old}}$ is the parameters for actor network before update. 
Accordingly, the parameters of critic are updated by minimizing 
\begin{IEEEeqnarray}{rCl}
  \mathcal{L}_c(\phi)=\mathbb{E}_{t}\left[\left( V \left(\mathbf{o}_t;\phi\right)-\hat{R}_t\right)^2\right].
\end{IEEEeqnarray}

The detail of the training is presented in Algorithm \ref{alg:RLwithPredict}. To begin with, the parameters for actor network and critic network are randomly initialized, while the parameters for GRUs are loaded as the trained model. During the interactions with environment, the prediction is first executed to generate the future information $x_t$; then the observation, i.e., the combination of current state $\mathbf{s}_t$ and $x_t$, is served as input to generate policy and estimated value; finally, the action is sampled according to the policy and executed. After collecting a batch of interactions, the algorithm would calculate the advantage $\hat{A}_t$ and reward-to-go $\hat{R}_t$, and update the networks several times.

\begin{algorithm}
  \caption{PPO for scheme with prediction}
  \label{alg:RLwithPredict}
  \begin{algorithmic}
      \State Initialize parameters $\theta$ and $\phi$ for actor and critic 
      \State Load parameters $\varphi$ for GRUs
      \For{episode $=0$ to $N$}
        \State $h_0\leftarrow 0$
        \For{$t=0$ to $T$}
          \State Perform prediction $x_t,h_{t+1}=GRU(\mathbf{s}_t,h_t;\varphi)$
          \State $\mathbf{o}_t\leftarrow[\mathbf{s}_t,x_t]$
          \State $\mathcal{P}\leftarrow\pi(\cdot|\mathbf{o}_t;\theta)$, $v_t=V(\mathbf{o}_t;\phi)$
          \State Sample action $a_t$ according to distribution $\mathcal{P}$
          \State Execute action $a_t$ and observe $\mathbf{s}_{t+1}$
          \State Compute the probability $p^{old}_t\leftarrow\mathcal{P}(a_t)$ 
        \EndFor
        \State $\hat{A}_T\leftarrow 0,v_T\leftarrow 0$ 
        \For{$t=T-1$ to $0$}
          \State $\hat{R}_t\leftarrow \gamma\lambda\hat{A}_{t+1}+r_t+\gamma v_{t+1}$
          \State $\hat{A}_t\leftarrow \hat{R}_t-v_t$
        \EndFor
        \For{$k=1$ to $K$}
          \State $\mathcal{L}_a\leftarrow 0, \mathcal{L}_c\leftarrow 0$
          \For{$t=0$ to $T-1$}
            \State $w_t\leftarrow \pi(a_t|\mathbf{o}_t;\theta)/p^{old}_t$
            \State $\mathcal{L}_a += \min \left(w_t\hat{A}_t,\text{clip}\left(w_t,1-\epsilon,1+\epsilon\right)\hat{A}_t\right)$
            \State $\mathcal{L}_c += \left(V(\mathbf{o}_t;\phi)-\hat{R}_t\right)^2$
          \EndFor
          \State Update $\theta$ and $\phi$ with gradient $\nabla_{\theta} \mathcal{L}_a$ and $\nabla_{\phi} \mathcal{L}_c$          
        \EndFor
      \EndFor
  \end{algorithmic}
\end{algorithm}

\subsection{DRL-Based Scheme without Prediction}
As shown in Fig. \ref{fig:network}, under the scheme with prediction, the RNN and MLP are trained with SL and RL, respectively. 
While the scheme without prediction performs end-to-end training for the whole networks. Put differently, the networks of actor and critic are comprised of RNN and MLP, rather than only MLP.
In this sense, during the training process, the RNN could automatically learn appropriate parameters such that the most important information at previous time steps could be captured for decision.

The detail of DRL-based scheme without prediction is express in Algorithm \ref{alg:RLwithoutPredict}. The distinctions between these two algorithms can be summarized as follows. First, the state $s_t$ is directly served as input to generate the policy and estimated value function.
Simultaneously, the hidden states for actor and critic are generated for the next calculation. Second, the parameters of GRU are updated during the RL training, rather than SL.

\begin{algorithm}
  \caption{PPO for scheme without prediction}
  \label{alg:RLwithoutPredict}
  \begin{algorithmic}
      \State Initialize parameter $\theta$ and $\phi$ for actor and critic 
      \For{episode $=0$ to $N$}
        \State $h_0^{\pi}\leftarrow 0, h_0^{V}\leftarrow 0$
        \For{$t=0$ to $T$}
          \State $\mathcal{P},h^{\pi}_{t+1}=\pi(\mathbf{s}_t,h_t^{\pi};\theta)$
          \State Sample action $a_t$ according to distribution $\mathcal{P}$
          \State $p^{old}_t\leftarrow \mathcal{P}(a_t)$ 
          \State $v_t,h^V_{t+1}=V(\mathbf{s}_t,h^V_{t};\phi)$
          \State Execute action $a_t$ and observe $\mathbf{s}_{t+1}$
        \EndFor
        \State $\hat{A}_T\leftarrow 0,v_T\leftarrow 0$ 
        \For{$t=T-1$ to $0$}
          \State $\hat{R}_t\leftarrow \gamma\lambda\hat{A}_{t+1}+r_t+\gamma v_{t+1}$
          \State $\hat{A}_t\leftarrow \hat{R}_t-v_t$
        \EndFor
        \For{$k=1$ to $K$}
          \State $\mathcal{L}_a\leftarrow 0, \mathcal{L}_c\leftarrow 0, h_0^{\pi}\leftarrow 0, h_0^{V}\leftarrow 0$
          \For{$t=0$ to $T-1$}
            \State $\mathcal{P},h^{\pi}_{t+1}=\pi(\mathbf{s}_t,h_t^{\pi};\theta)$
            \State $V_t,h^{V}_{t+1}=V(\mathbf{s}_t,h_t^{V};\phi)$
            \State $w_t\leftarrow \mathcal{P}(a_t)/p^{old}_t$
            \State $\mathcal{L}_a += \min \left(w_t\hat{A}_t,\text{clip}\left(w_t,1-\epsilon,1+\epsilon\right)\hat{A}_t\right)$
            \State $\mathcal{L}_c += (V_t-\hat{R}_t)^2$
          \EndFor
          \State Update $\theta$ and $\phi$ with gradient $\nabla_{\theta} \mathcal{L}_a$ and $\nabla_{\phi} \mathcal{L}_c$          
        \EndFor
      \EndFor
  \end{algorithmic}
\end{algorithm}

\section{Performance Evaluation}
In this section, the performances of DRL-based energy management scheme with and without prediction 
are compared. First, the simulation environment settings and algorithmic implementation are provided.
Then, the performances of prediction used for energy management scheme are evaluated.
Finally, the simulation results of comparisons between two schemes and corresponding explanation are given.

\subsection{Environment Setup}
We consider the energy management problem during one day, such that the time horizon $T$ is 24.
The transition functions and cost function are specified as follows \cite{TSGQin}.
\begin{IEEEeqnarray}{rCl}
  f\left(b, a\right) = 
  \begin{cases}
    b  + \frac{\eta_{c}}{C}a,&\mbox{if } a\geq 0,\\
    b + \frac{1}{C\eta_{d}}a,&\mbox{otherwise},\\
  \end{cases}\\
  g\left(b, a\right) = 
  \begin{cases}
    \lambda_1|a|,&\mbox{if } b< 0.5,\\
    \lambda_2|a|,&\mbox{otherwise},\\
  \end{cases}
\end{IEEEeqnarray}
where $C$, $\eta_{c}$ and $\eta_{d}$  denotes the capacity, charging and discharging efficiency coefficients of BESs,
$\lambda_1$ and $\lambda_2$ are the maximum and minimum degradation
cost per kWh, corresponding to low SOC and high SOC, respectively. 
The parameter settings are provided in Table \ref{table:parameter}.

\begin{table}[htbp]
  \scriptsize
  \centering
  \caption{Environment and Algorithmic Parameter Settings}    \label{table:parameter}
  \begin{tabular}{cccccc}
    \toprule
    \textbf{Parameter} & \textbf{Value} & \textbf{Parameter} & \textbf{Value} & \textbf{Parameter} & \textbf{Value}\\
    \midrule
     $d_{max}$& 400 kW & $c_{max}$ & 400 kW & $C$ & 2000 kWh \\
    $\eta_d$  & 0.95 & $\eta_c$ & 0.95 & $\lambda_1$ & 0.013  \\
    $\lambda_2$ & 0.005 & $\epsilon$ & 0.2 & $K$ & 3 \\
    \bottomrule
  \end{tabular}
\end{table}

We employ 10 parallel threads to interact with the environment.
We use the historical data from 2015-01-05 to 2018-12-17 for the SL training, 
while the data from 2018-12-18 to 2020-03-23 is used for the test of prediction performance.
For the training of DRL, the discount factor $\gamma$ is set to be 0.95, 
The learning rate of actor and critic is set to be $3\times 10^{-4}$ and $1\times 10^{-3}$, respectively.
Other important algorithmic parameters are shown in Table \ref{table:parameter}.

\subsection{Performance of Prediction}
\begin{figure}[htbp]
  \centering{\includegraphics[width=.4\textwidth]{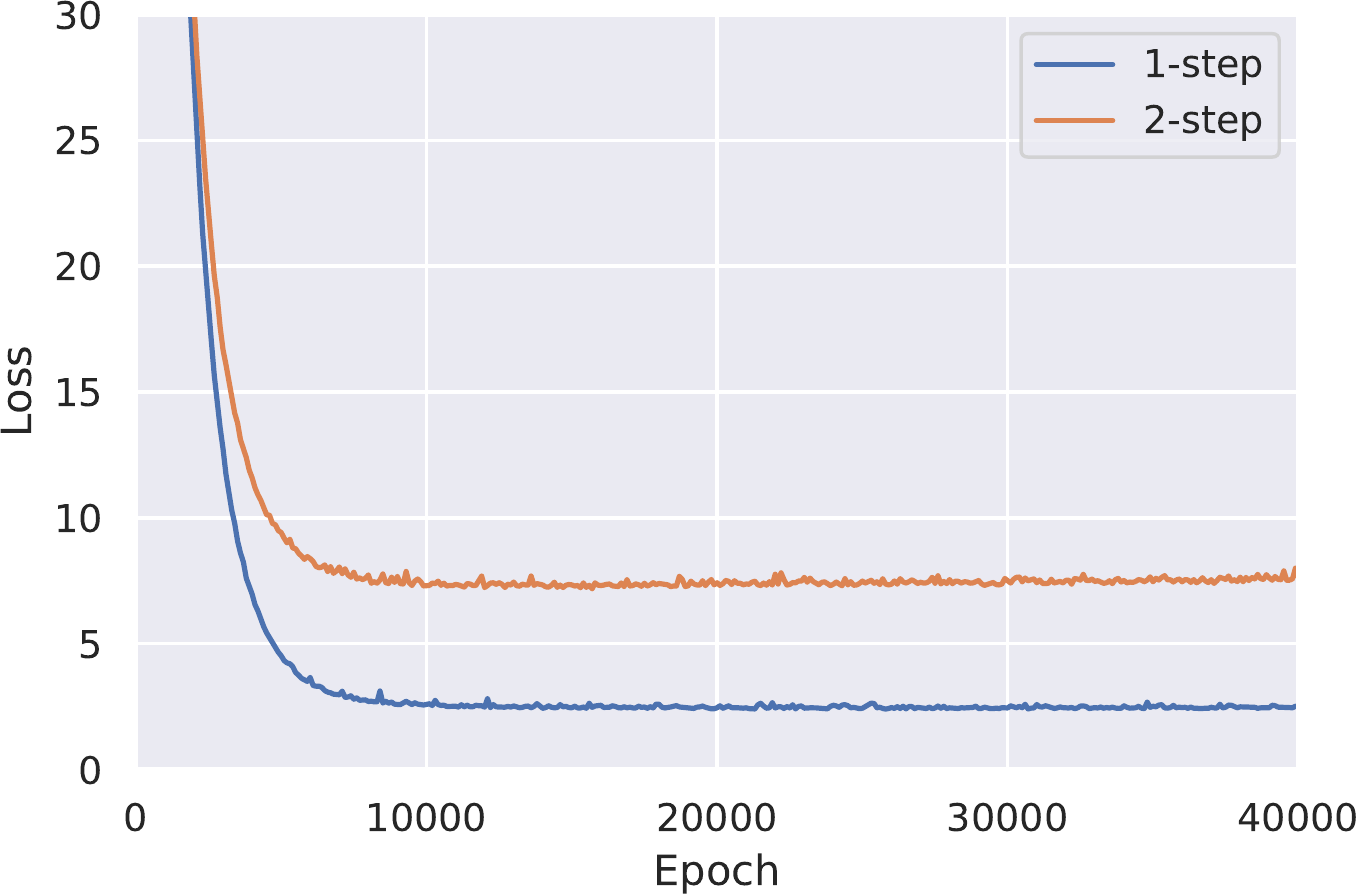}}
  \caption{Loss curve of prediction for renewable generation output.}
  \label{fig:gen_loss}
\end{figure}

\begin{figure}[htbp]
  \centering{\includegraphics[width=.4\textwidth]{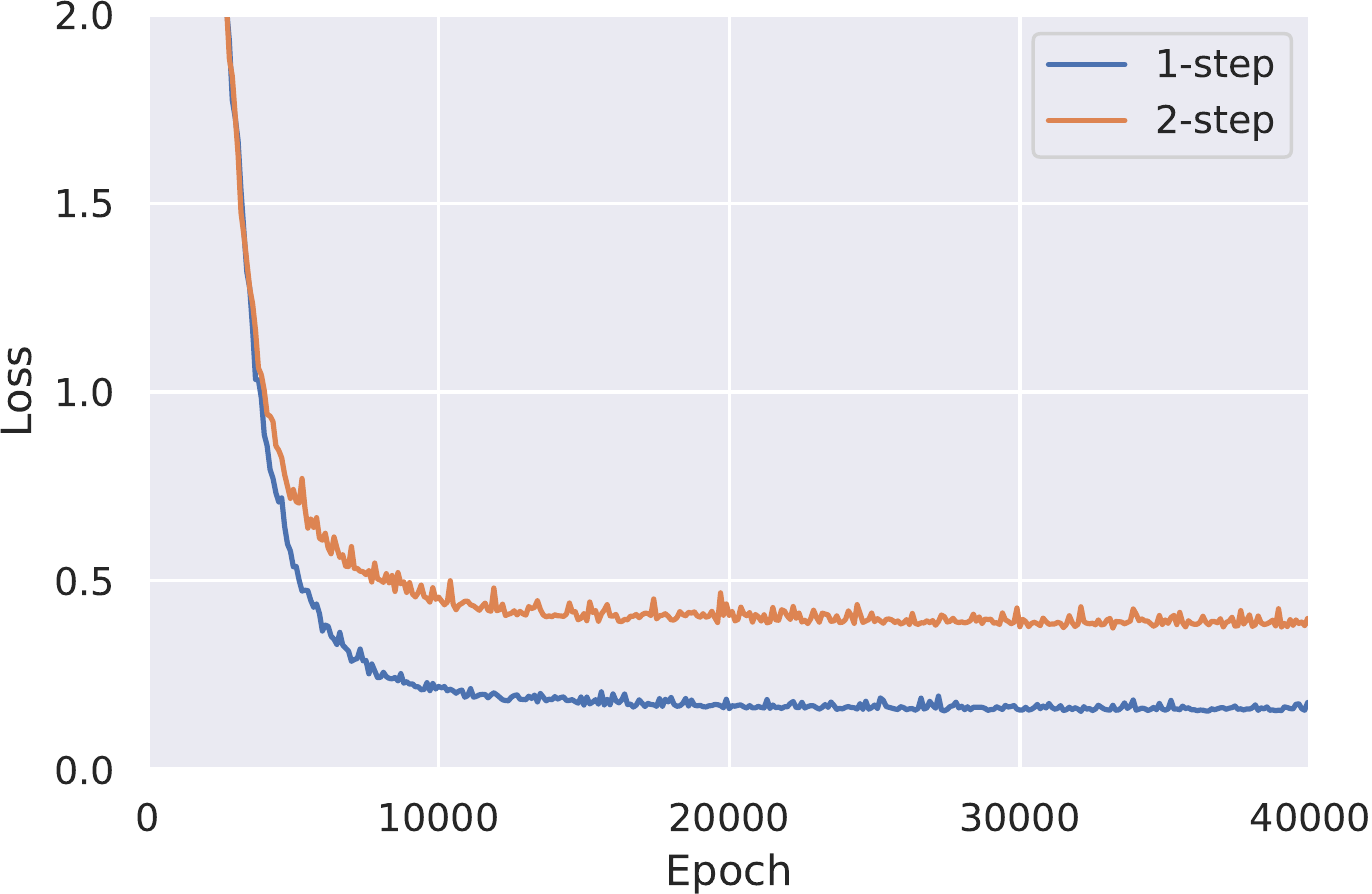}}
  \caption{Loss curve of prediction for power demand.}
  \label{fig:load_loss}
\end{figure}

\begin{figure}[htbp]
  \centering{\includegraphics[width=.4\textwidth]{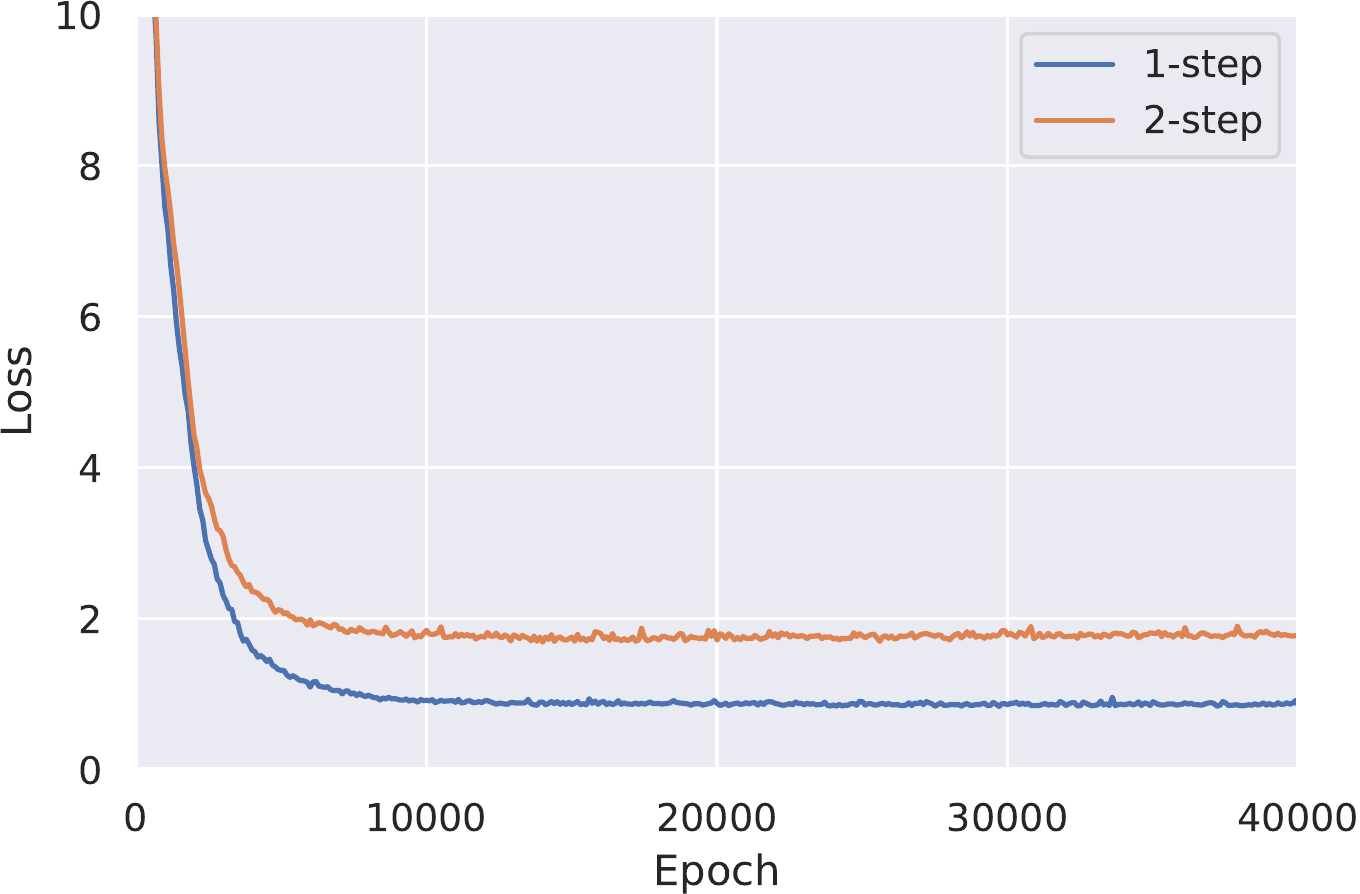}}
  \caption{Loss curve of prediction for electricity price.}
  \label{fig:price_loss}
\end{figure}

During training process of SL, The MSE losses of the prediction for renewable generation output, 
power demand and electricity price are shown in Fig. \ref{fig:gen_loss}, Fig. \ref{fig:load_loss} and Fig. \ref{fig:price_loss}, respectively.
It can be observed from these three figures that the MSE losses rapidly decrease and eventually converge, 
which demonstrates the stable training of RNNs.

\begin{figure}[htbp]
  \centering{\includegraphics[width=.4\textwidth]{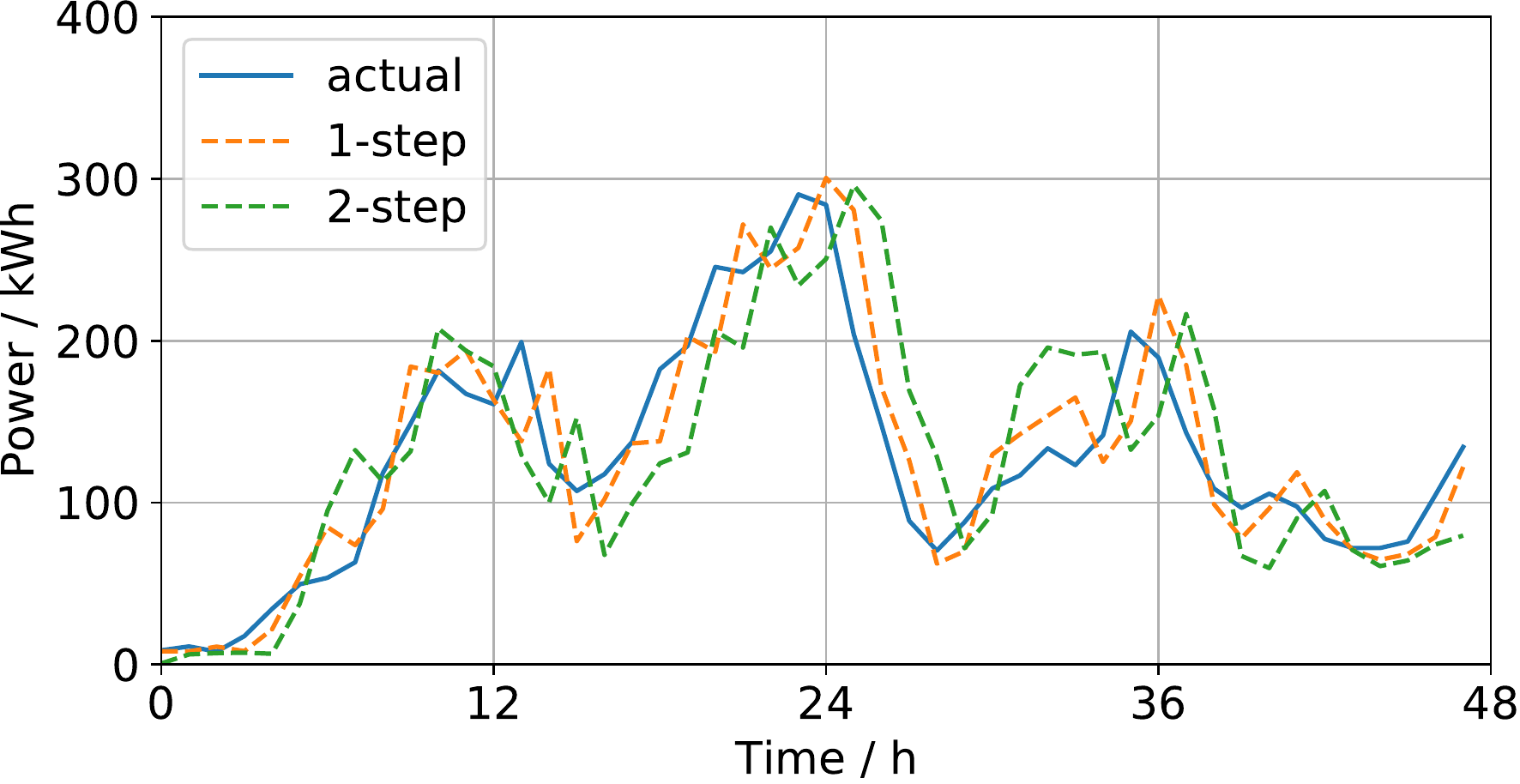}}
  \caption{Prediction for renewable generation output.}
  \label{fig:gen_predit}
\end{figure}

\begin{figure}[htbp]
  \centering{\includegraphics[width=.4\textwidth]{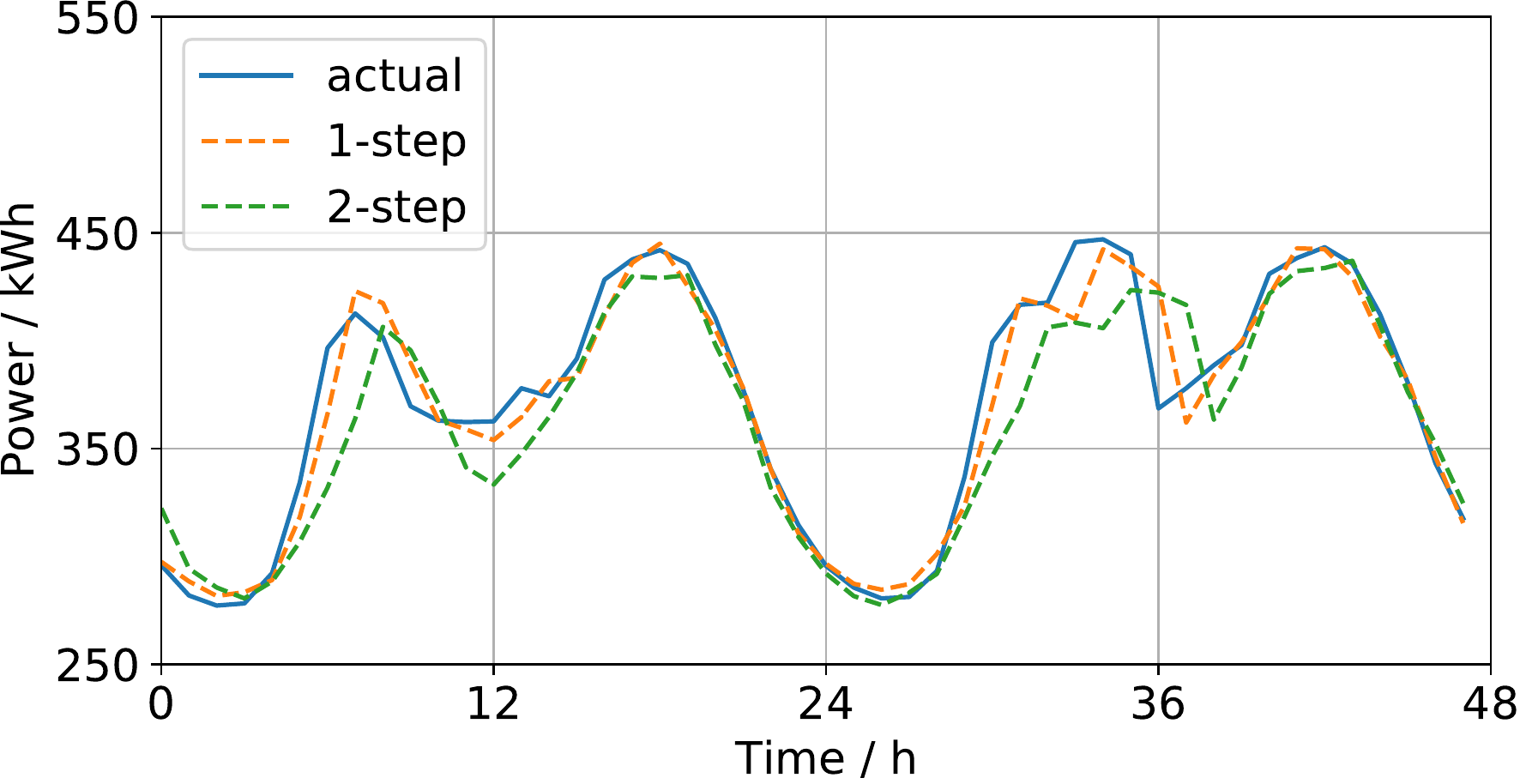}}
  \caption{Prediction for power demand.}
  \label{fig:load_predit}
\end{figure}

\begin{figure}[htbp]
  \centering{\includegraphics[width=.4\textwidth]{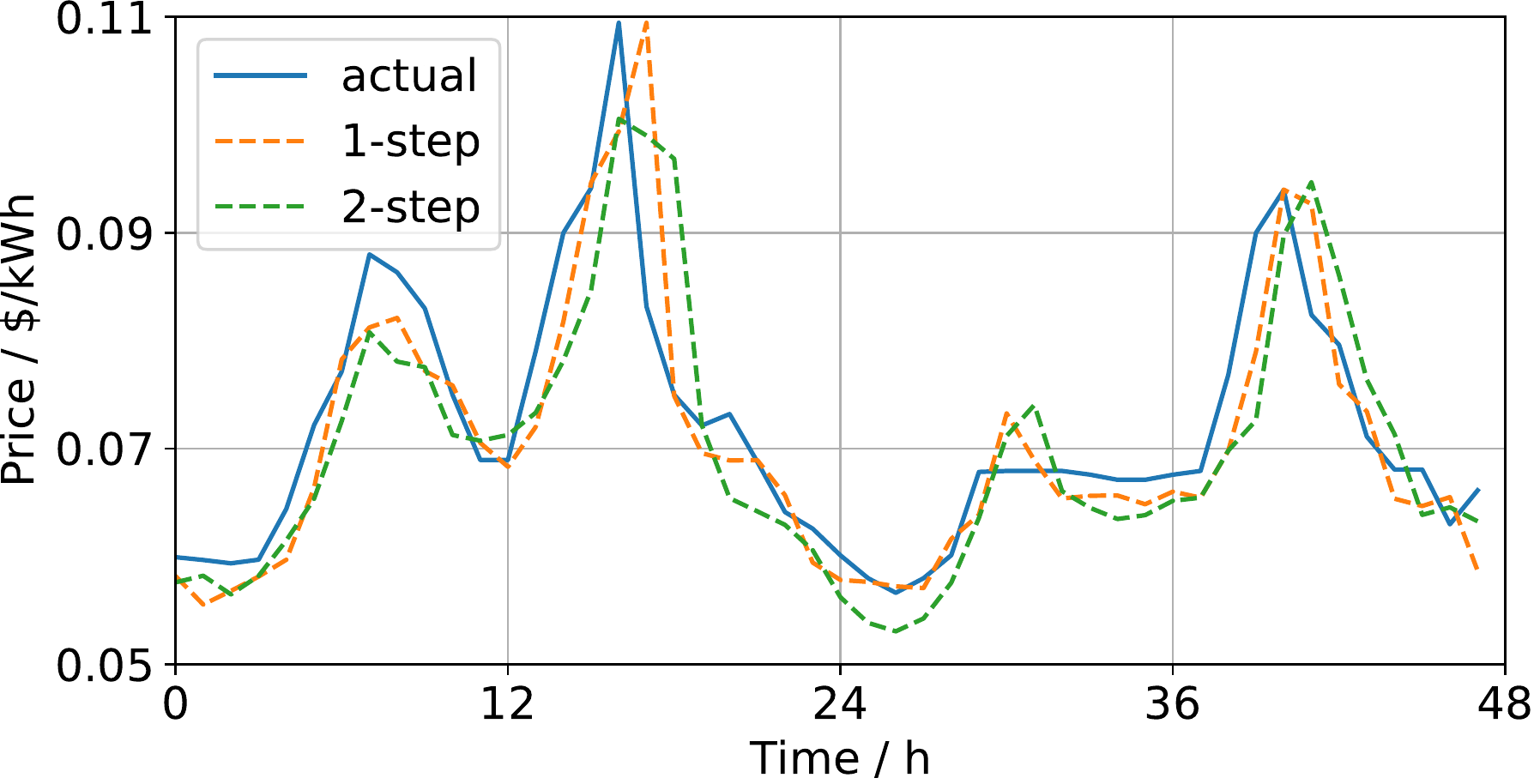}}
  \caption{Prediction for electricity price.}
  \label{fig:price_predit}
\end{figure}

The prediction effects during two days are shown in Fig. \ref{fig:gen_predit}, Fig. \ref{fig:load_predit} and Fig. \ref{fig:price_predit}.
It can be observed from these three figures that the 1-step prediction is more accurate than 2-step prediction, which is also revealed by the loss curves during training. 

We adopt two metrics to evaluate the performances of prediction: mean absolute percentage error (MAPE) and root-mean-square error (RMSE).
The evaluation results are shown in Table \ref{table:predit}. We can observe that the statistic results are consistent with the intuitive illustration in Fig. \ref{fig:gen_predit}-\ref{fig:price_predit}: 1-step prediction is more precise than 2-step prediction.

\begin{table}[htbp]
  \scriptsize
  \centering
  \caption{Test Performance for Prediction}    \label{table:predit}
  \begin{tabular}{ccc|cc}
    \toprule
     & \multicolumn{2}{c|}{\textbf{1-step}} & \multicolumn{2}{c}{\textbf{2-step}}\\
    \midrule
    \textbf{Metrics} & MAPE & RMSE & MAPE & RMSE\\
    \midrule
    Renewable generation output & $17.7\%$ & 31.0 & $31.0\%$ & 47.0  \\
    Power demand  & $3.0\%$  & 13.3 & $5.7\%$ & 22.8 \\
    Electricity Price & $8.2\%$ & 0.0046 &  $11.4\%$ & 0.0058 \\
    \bottomrule
  \end{tabular}
\end{table}

\subsection{Performance of Energy Management Schemes}
\begin{figure}[htbp]
  \centering{\includegraphics[width=.45\textwidth]{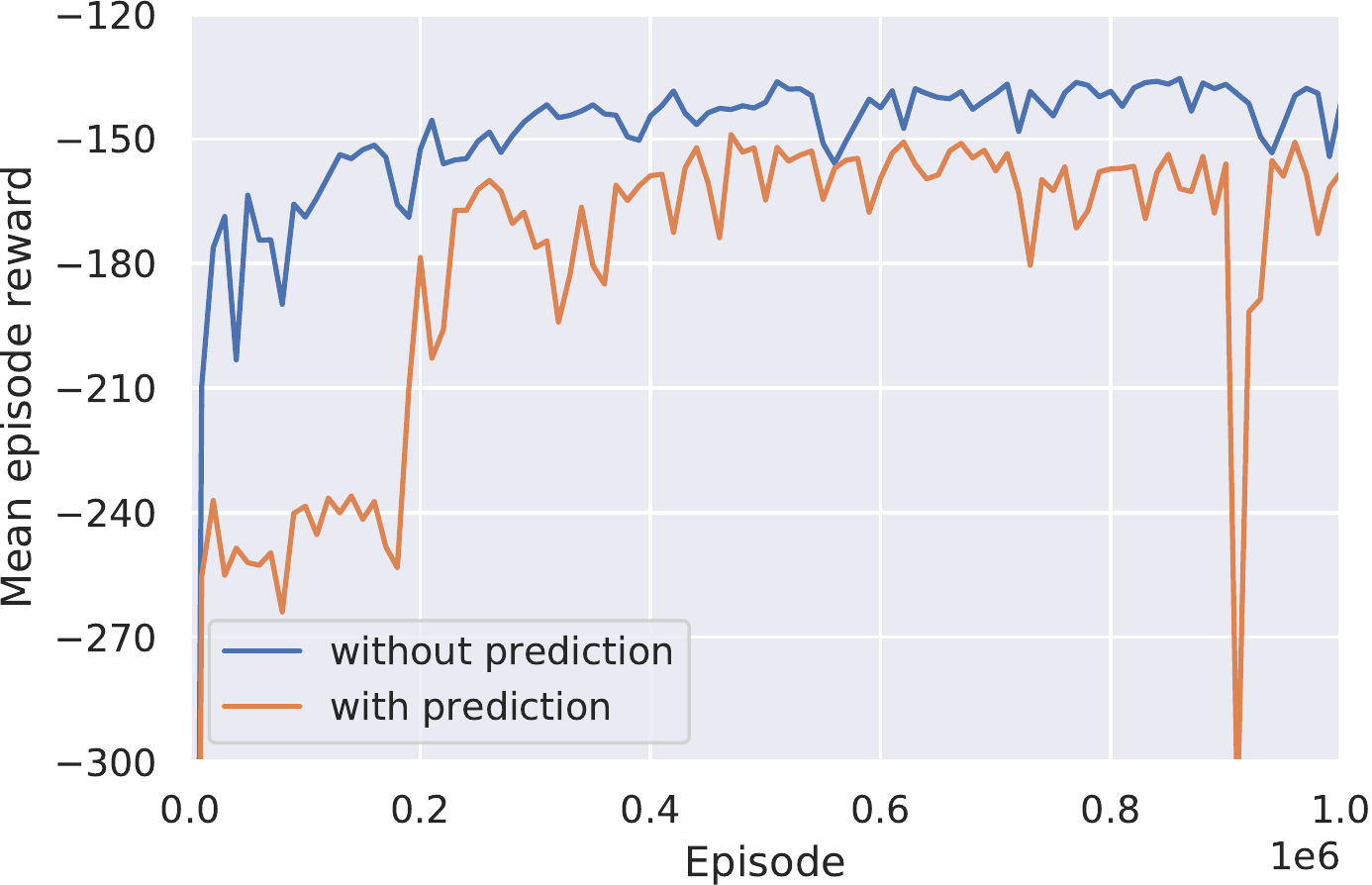}}
  \caption{Training curve.}
  \label{fig:train}
\end{figure}

The evaluate the performances of these two schemes during the training process of PPO, 
we depict the mean episode reward ($R=\sum_{t=0}^T r_t$) in Fig. \ref{fig:train}.
We can see that the energy management scheme without prediction has higher episode reward than scheme with prediction. Moreover, the performance of the scheme without prediction is rapidly promoted to a relatively high level (more than -180), while the counterpart achieves the same performance after $2\times10^5$ episodes.

\begin{figure}[htbp]
  \centering{\includegraphics[width=.45\textwidth]{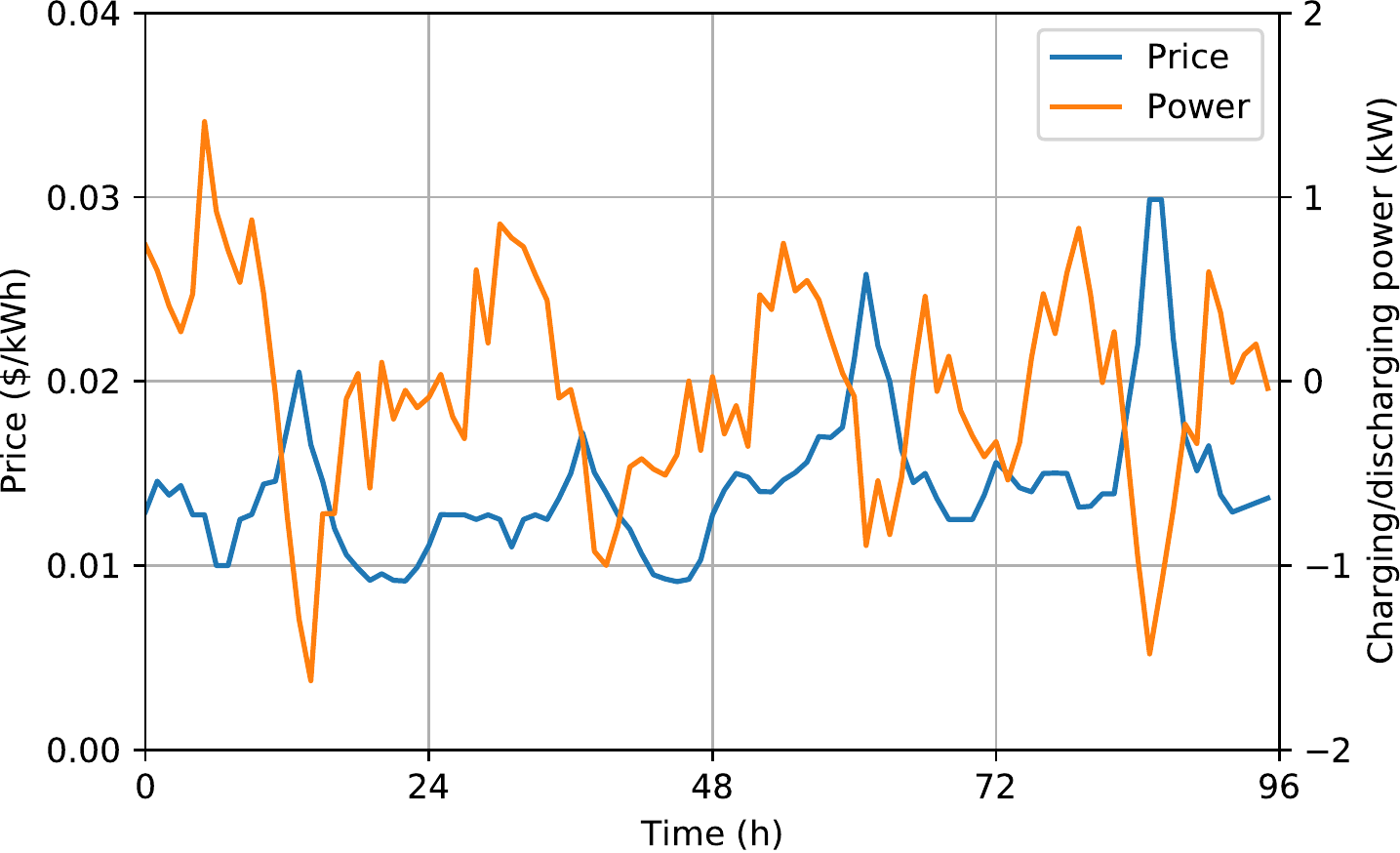}}
  \caption{Charging/discharging power of the scheme without explicit prediction over 4 consecutive days.}
  \label{fig:power}
\end{figure}

Under the energy management scheme without prediction,
the curves of charging/discharging power of BESs with electricity price are shown in Fig. \ref{fig:power}. 
One can observe that the scheme without explicit prediction learns to charge when the electricity
price is low and to discharge when the price is on-peak. These
charging/discharging patterns demonstrate the DRL-based energy management scheme can accommodates the varying electricity price without explicit prediction.

\section{Conclusion}
In this paper, we investigate whether the prediction matters in DRL-based energy management scheme.
We present the standard energy management scheme with and without explicit prediction.
The former is implemented with both SL and DRL, while the latter is directly implemented with end-to-end DRL.
The simulation results demonstrate that end-to-end DRL enables the EMS to learn better control policies without explicit prediction sessions.
This work can clarify the misuse and misunderstanding for DRL methods in the field of energy management.

\bibliographystyle{IEEEtran}
\bibliography{mybibfile}

\end{document}